\documentclass{pasa}%

\title[Mode identification of HD~189631]{Spectroscopic Pulsational Frequency
and Mode Determination of the $\gamma$~Doradus Star HD\,189631}
\author[M. W. Davie et al.]{M. W. Davie$^1$\thanks{m.davie@student.unsw.edu.au}, K. R. Pollard$^2$, P. L. Cottrell$^2$, E. Brunsden$^3$, D. J. Wright$^1$, P. De Cat$^4$ \\
\affil{$^1$Department of Astrophysics, University of New South Wales, Sydney, NSW 2052, Australia}
\affil{$^2$Department of Physics and Astronomy, University of Canterbury, Private Bag 4800, Christchurch, New Zealand}
\affil{$^3$Department of Physics, University of York, Heslington, York, YO10 5DD, UK}
\affil{$^4$Royal Observatory of Belgium, Ringlaan 3, 1180 Brussels, Belgium}}

\jid{PASA}
\doi{10.1017/pas.\the\year.xxx}
\jyear{\the\year}

\usepackage[authoryear]{natbib}
\bibpunct{(}{)}{;}{a}{}{,}
\newcommand{\aap}{A\&A}

\begin{document}%
\begin{abstract}
We present improvement and confirmation of identified frequencies and pulsation modes for the $\gamma$~Doradus star HD~189631. This work improves upon previous studies by incorporating a significant number of additional spectra and precise determination of frequencies. Four frequencies were identified for this star: $1.6774 \pm 0.0002$ d$^{-1}$, $1.4174 \pm 0.0002$ d$^{-1}$, 
$0.0714 \pm 0.0002$ d$^{-1}$, and $1.8228 \pm 0.0002$ d$^{-1}$ which were identified with the modes ($l$,$m$) = ($1,+1$), ($1,+1$), ($2,-2$), and ($1,+1$) respectively. These findings are in agreement with the most recent literature. The prevalence of ($l$,$m$) = ($1,+1$) modes in $\gamma$~Doradus stars is starting to become apparent and we discuss this result.
\end{abstract}
\maketitle%
\section{INTRODUCTION }
\label{sec:intro}

Asteroseismology is a mature field of study and the associated concepts and methodologies have been established for some time. 
The use of stellar pulsations as a probe of stellar structure is recognised as a powerful tool in building understanding of  the interior structure and evolution of stars. The gravity modes (g-modes) characteristic of the $\gamma$~Doradus
stars are deeply penetrative and so provide a way to characterise the deep stellar interior and layers close to the core. Accurately finding and characterising these g-modes is the first step in obtaining a more complete understanding of the $\gamma$~Doradus class of pulsating stars.

This work improves upon a prior set of spectroscopically identified modes in \cite{2011MNRAS.415.2977M} and \citet{Tkachenko2013} by incorporating new spectra from other sites, providing a longer timebase of observations to improve frequency determination, and incorporating a more detailed exploration of the mode-identification parameter-space.

HD~189631 was selected for study as it displays robust line-profile variation. \citet{Perryman1997} classifies HD 189631 
as an A9V star and Vizier \citep{Ochsenbein2000} lists the star as an F0V. The star has a relatively bright visual magnitude of 7.54 making it a good target for high-resolution spectroscopic observation and analysis.
HD~189631 has been found to have a moderate \emph{v}sin\emph{i} = 43.6 $\pm$ 0.5 kms$^{-1}$ and a mean radial velocity of --10.13 kms$^{-1}$ \citep{2011MNRAS.415.2977M}.

\section{Observations and data treatment}\label{obs}

In total just over 100 spectra of HD 189631 were collected from the 1-metre
McLellan telescope at Mt John University Observatory (MJUO) in Tekapo, New
Zealand with the fibre-fed High Efficiency and
Resolution Canterbury University Large Echelle Spectrograph (HERCULES). Spectra from MJUO were collected over a period of $3$ years from $2008$~July to $2011$~ June. The main improvement of this paper over  that of  \citet{2011MNRAS.415.2977M} is the incorporation of 58 new spectra, increasing the timebase over which spectra were acquired from 411 days to 1118 days and thereby improving the precision of the resultant frequencies. These spectra were reduced using a customisable \textsc{matlab} pipeline initially written by Dr. Duncan Wright \citep{2012MNRAS.tmp.2786B} and barycentric corrections applied.

Spectra were also obtained during  $2008$~July -- August  and $2009$~June -- July using HARPS and FEROS at the ESO in La Silla, Chile (see  Table \ref{189631obstable}). A total of $376$ spectra were obtained at the ESO.
These spectra are dominated by the very high-cadence observational campaign in $2008$ at La Silla Chile (Table \ref{189631obstable}). The spectra from both La Silla spectrographs were initially reduced on-site using an automated pipeline. Once reduced, each of the spectra were then cross-correlated against a delta-function comb as described in \citet{2012MNRAS.tmp.2786B}  to
create a representative line-profile for each of the spectra.

\begin{table}
\begin{minipage}{0.45\textwidth}
\centering
\caption{Spectra of HD 189631 collected for analysis in this project. 
The lower section of the table displays spectra that are new to this work. Julian Dates given here are JD-2450000}
\label{189631obstable}
\begin{tabular}{cccr}
\hline \hline
JD&Observer&Instrument&Obs\\
\hline
4620-4626&P.~De~Cat&HARPS\footnote{ESO, La Silla, Chile.}&272\\
4651-4662&P.~Kilmartin&HERCULES\footnote{MJUO, Lake Tekapo, New Zealand.}&38\\
4660-4667&L.~Mantegazza&FEROS\footnote{ESO, La Silla, Chile.}&56\\
4683-4685&K.~Pollard&HERCULES&7\\
5003-5012&E.~Poretti&HARPS&23\\
5028-5031&J.C.~Su\'{a}rez&HARPS&25\\
\hline
5363&P.~Kilmartin&HERCULES&1\\
5455-5458&P.~Kilmartin&HERCULES&4\\
5703-5712&E.~Brunsden&HERCULES&28\\
5720-5738&P.~Kilmartin&HERCULES&25\\
\hline
Total&&&479\\
\hline\hline
\end{tabular}
\end {minipage}
\end{table}

Frequency analyses were undertaken using \textsc{sigspec} \citep{2007A&A...467.1353R} and \textsc{famias} \citep{2008CoAst.155...17Z} and the subsequent
mode identification used \textsc{famias}. Frequecies were analysed both from moments as in \citet{2003A&A...398..687B} and pixel-by-pixel across the line profile.
Mode identification in \textsc{famias} uses the Fourier Parameter Fit method; for each identified frequency, synthetic spectra for different modes 
are generated from a range of parameters until the fit is optimised by minimisation of the $\chi^{2}$ parameter.

The previous spectroscopic frequencies 
and modes identified for HD 189631 from \citet{2011MNRAS.415.2977M} and from \citet{Tkachenko2013} may be found in Table \ref{189631prior}.

\begin{table}

\caption[Literature results for HD 189631 frequency and mode identification]{Previously determined frequencies and modes for HD 189631 from \citet{2011MNRAS.415.2977M} in the second and third columns, and \citet{Tkachenko2013} in the fourth and fifth columns.}
\label{189631prior}
\begin{minipage}{0.45\textwidth}
 \centering
 \begin{tabular}{lcccc}  
 \hline \hline
 &f (d$^{-1}$)& Mode ($l$,$m$)&f (d$^{-1}$)& Mode ($l$,$m$)\\
  \hline
 \textit{f}$_{1}$ &$1.6719$& ($1,+1$)&$1.685$&($1,+1$)\\
 \textit{f}$_{2}$ &$1.4200$& ($3,-2$)&$1.411$&($1,+1$)\\
 \textit{f}$_{3}$ &$0.0711$& ($2,-2$)&$0.122$&---\\
 \textit{f}$_{4}$ &$1.8227$& ($4,+1$)\footnote{The fit of this mode was uncertain in \citet{2011MNRAS.415.2977M}. It was suggested that a ($l$,$m$) = ($2,-2$) mode may also be a good candidate for this frequency.}&$1.826$&($1,+1$)\\
   \hline \hline
\end{tabular}\par
   \vspace{-0.75\skip\footins}
   \renewcommand{\footnoterule}{}
\end {minipage}
\end{table}

\begin{figure}
 \includegraphics[width=0.45\textwidth,keepaspectratio=true]{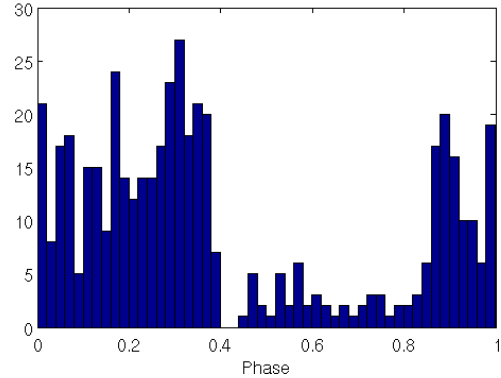}

 \caption[1-day phase sampling of spectra for HD 189631]{Spectra for HD 189631 phased to $1$ d$^{-1}$ showing uneven sampling of spectra.}
 \label{189631_1dayphase}
\end{figure}

In this work additional spectra were obtained, substantially broadening the timebase (from $411$ d to $1118$ d, see Table \ref{189631obstable}) for determination of frequencies. The key 
outcome of this is to improve the frequency determination of the stellar pulsational behaviour and the confidence in the subsequent modes identified for them. 
The spectra obtained for this study were from two sites. Whilst this does offer some relief from the 1-day sampling problem it does not alleviate it completely as 
there are many more spectra obtained from HARPS and FEROS than from HERCULES. Figure \ref{189631_1dayphase} shows the uneven distribution of spectra over a period of time phased on one day.

In combining the cross-correlated line profiles from the three spectrographs (HARPS, 
FEROS and HERCULES) into one data set, the mean line profiles needed to be 
similar. As HARPS has the highest resolution of the three spectrographs, a greater number of lines can be chosen 
for creating the $\delta$-function mask for the set by using the HARPS spectra. The cross-correlated line profiles 
from FEROS and HERCULES were scaled such that their mean line profiles were similar to the mean line profile from HARPS.

\section{Frequency Analysis}\label{frequency}

\begin{figure}

 \includegraphics[width=0.45\textwidth,keepaspectratio=true]
 {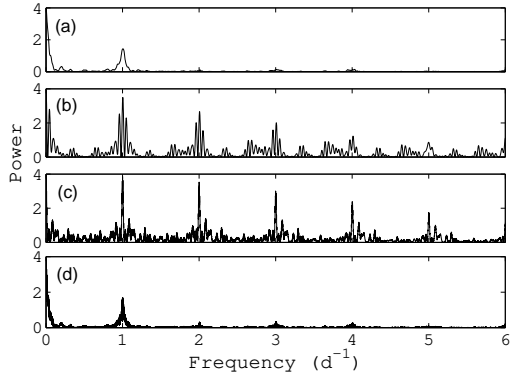}

 \caption[HD 189631 spectral window by site] {Spectral window for HD 189631 observations per site. (A) HARPS; (B) FEROS, (C) HERCULES, (D) all observations combined.}
 \label{189631window}
\end{figure}

Figure \ref{189631window} shows the spectral windows for observations of HD~189631 from each contributing site and as a whole. Distinctive peaks are 
readily seen at one cycle-per-day and subsequent aliases. Figure \ref{189631window}a shows the spectral window for the HARPS observations. There 
are $273$ observations over a relatively short timeframe. That they were collected over just a few weeks is significant, as the peaks in the spectrum are broad and 
give relatively poor definition of the frequencies. In comparison, the other spectra were taken over a longer timespan, but with fewer spectra in total.

The Fourier spectra from FEROS appear particularly noisy owing to the 
relatively low number of observations and short timebase over which the observations span. In comparison, the spectra from HERCULES are 
of a similar number but over a much longer timebase giving better resolution of the peak and more depressed side-lobes than FEROS, 
but a similar level of 1-day aliasing. 

The number of HARPS spectra means that it dominates in the collective spectral window (Figure \ref{189631window}d), but the 
collective timebase is now much longer, giving good sharpness of peaks for frequency determination. 

The frequencies and the significance of each identified frequency were calculated using \textsc{sigspec} \citep{2011CoAst.163....3R}.
An conservative acceptable significance for an identified frequency was deemed to be $sig(f) > 15$ following \citet{2012MNRAS.427.2512B}.

\subsection{Frequency analysis from moments}

\begin{table*}
\centering
\caption[Frequency determination from analysis of moments for HD 189631]{Frequencies determined from analysis of moments for HD 189631. Frequencies for the zeroth ($M_0$) to third ($M_3$) moments are shown. The \textsc{sigspec} significance for each frequency is given by $sig(f)$.}
\label{189631momfreq}
\begin{minipage}{0.99\textwidth}
 \centering
 \begin{tabular}{l|cc|cc|cc|cc}  
 \hline \hline
  	&\multicolumn{2}{c|}{$M_0$}	&\multicolumn{2}{c|}{$M_1$}	&\multicolumn{2}{c|}{$M_2$}	&\multicolumn{2}{c}{$M_3$}	\\
\hline
	&\multicolumn{2}{c|}{Frequency}&\multicolumn{2}{c|}{Frequency}&\multicolumn{2}{c|}{Frequency}&\multicolumn{2}{c}{Frequency}	\\
	&d$^{-1}$	&	$sig(f)$&d$^{-1}$	&	$sig(f)$&d$^{-1}$	&	$sig(f)$&d$^{-1}$	&	$sig(f)$\\
\hline
$f_{m1}$	&	1.004	&	56	&	1.848	&	28	&	1.423	&	36	&	1.420	&	31	\\
$f_{m2}$	&	3.009	&	24	&	1.675	&	28	&	1.677	&	31	&	0.071	&	30	\\
$f_{m3}$	&	0.039	&	23	&	0.993	&	28	&	0.598	&	25	&	1.794	&	30	\\
$f_{m4}$	&	4.996	&	9	&	1.420	&	26	&	0.139	&	23	&	1.666	&	27	\\
$f_{m5}$	&		&		&	0.071	&	24	&	2.806	&	16	&	0.244	&	18	\\
$f_{m6}$	&		&		&	0.561	&	18	&	0.968	&	16	&	0.993	&	16	\\
$f_{m7}$	&		&		&	0.866	&	17	&	3.507	&	11	&	2.175	&	16	\\
$f_{m8}$	&		&		&	1.944	&	13	&	3.160	&	11	&	0.564	&	14	\\
 \hline \hline
\end{tabular}
\end {minipage}
\end{table*}

The first four moments were calculated from the cross-correlated line profiles in \textsc{famias}. \textsc{sigspec} was used to determine the frequencies and their significances, $sig(f)$, found in the zeroth to third moments of the cross-correlated line profiles (see Table 3). Strongly seen in the zeroth moment is the $1$d$^{-1}$ frequency ($f_{m1}$) and two aliases of it at $f_{m2}$ and $f_{m4}$. The $1$d$^{-1}$ frequency also appears as $f_{m3}$ in $M_1$ and as $f_{m6}$ in $M_3$.

In the first moment ($M_1$), several strong frequencies are noted that also appear in other 
moments. In $M_1$, $f_{m1}$ ($1.848$ d$^{-1}$) appears very strongly but only marginally visible in the other moments as perhaps a 1-day alias of $f_{m7}$ in $M_1$, $f_{m3}$ in $M_3$, and perhaps $f_{m5}$ in $M_2$. $f_{m2}$ at $1.675$ d$^{-1}$ appears in $M_2$ as $f_{m2}$ ($1.677$ d$^{-1}$) and again 
in $M_3$ as $f_{m4}$ (1.666 d$^{-1}$). This frequency corresponds to $f_1$ as found by \citet{2011MNRAS.415.2977M} (see Table \ref{189631prior}). $f_{m4}$ ($1.420$ d$^{-1}$) appears as $f_{m1}$ in both the second and third moments ($1.423$ d$^{-1}$ and $1.420$ d$^{-1}$ respectively). It may also appear in $M_1$ as $f_{m1}$; a possible combination of twice $f_{m4}$ - $f_{m3}$. This frequency corresponds to $f_2$ in \citet{2011MNRAS.415.2977M}. In $M_1$, $f_{m5}$ ($0.071$ d$^{-1}$) also appears as $f_{m2}$ in $M_3$ and as $f_3$ in \citet{2011MNRAS.415.2977M}. $f_{m6}$ ($0.561$ d$^{-1}$) in $M_1$ is also seen (for each moment) as $f_{m8}$ ($0.564$ d$^{-1}$) in $M_3$.)

\subsection{Frequencies from pixel-by-pixel analysis}

Frequencies were determined from a pixel-by-pixel analysis of the cross-correlated line profiles. Fourier spectra were created and the 
frequency corresponding to the highest peak was chosen if above the significance level 
\citep{2008CoAst.155...17Z}. A pre-whitening was carried out, removing this . A pre-whitening was carried out, removing this 
frequency and the residuals used as the basis for the next frequency determination.

The determination of frequencies by \textsc{famias} is perhaps optimistic, as the calculation of the noise floor gives a rather conservative bound on the distinction of noise from a real signal. This may 
omit significant frequencies if taken alone. Using \textsc{SigSpec} also has problems in that it can over-detect frequencies as it takes little 
consideration of the intrinsic uncertainty from the contributing spectra. These two techniques were used together to 
obtain a reasoned set of frequencies for analysis.

The retention of the frequencies for further analysis was decided on the basis that it appears strongly in both the pixel-by-pixel analysis and the first moment Fourier spectra.
The uncertainties derived from $sig(f)$ in Table \ref{189631momfreq} using the equation of \citet{2008A&A...481..571K} were used with the final set of frequencies (Table \ref{189631finalfreq}).

\begin{figure}
  \centering
  \caption[Phase sampling for the HD 189631 observations]{Phase sampling for the four frequencies over the dates of observations for HD 189631. (a) $f_1$, (b) $f_2$, (c) $f_3$, (d) $f_4$.~$f_3$ has the least well sampled phase which is to be expected as the observed period is of the order of two weeks.}
  \label{189631_phasesampling}
\renewcommand{\tabcolsep}{1pt}
  \begin{tabular}{cc}

    \includegraphics[width=0.23\textwidth,keepaspectratio=true]{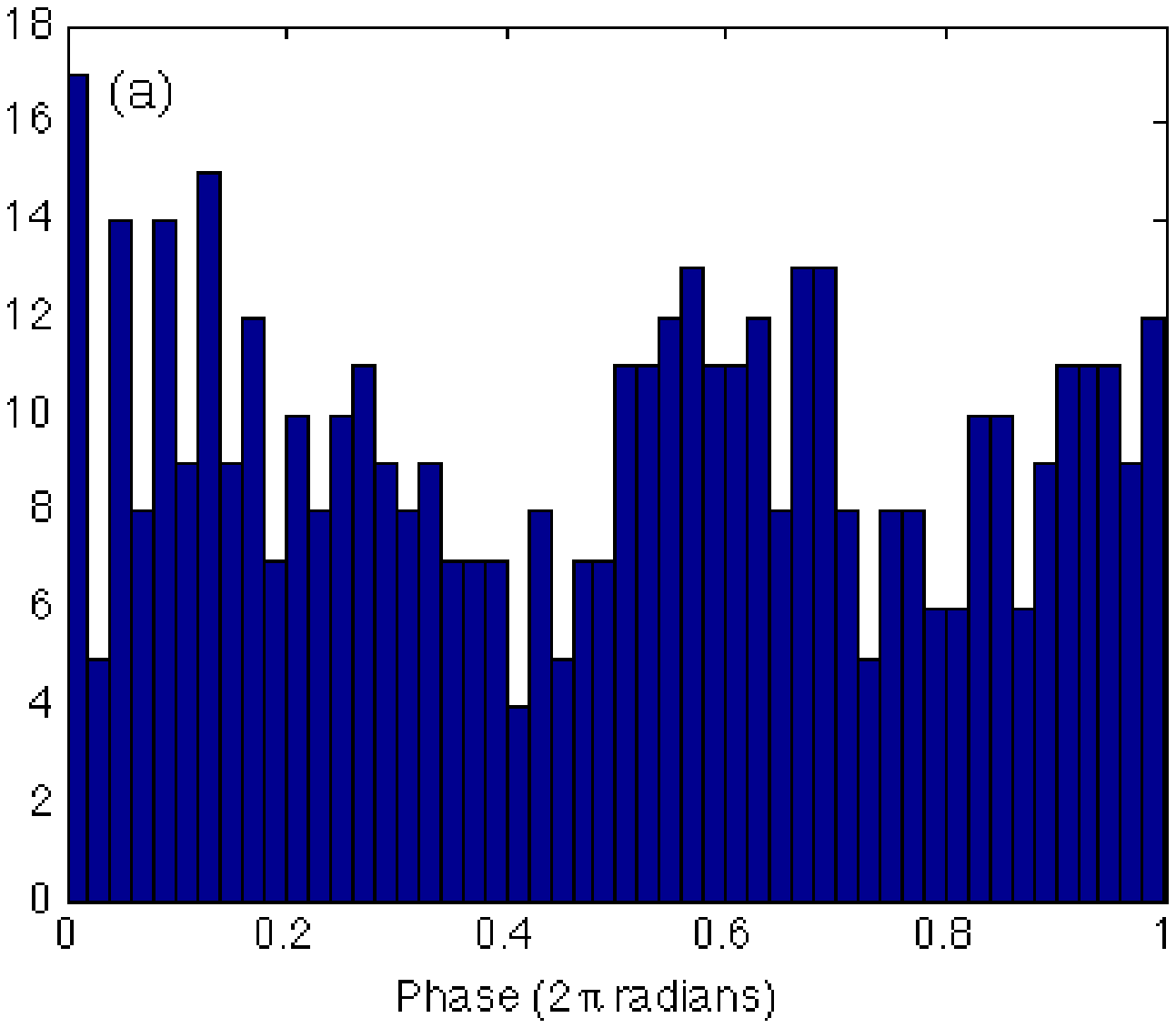}&

    \includegraphics[width=0.23\textwidth,keepaspectratio=true]{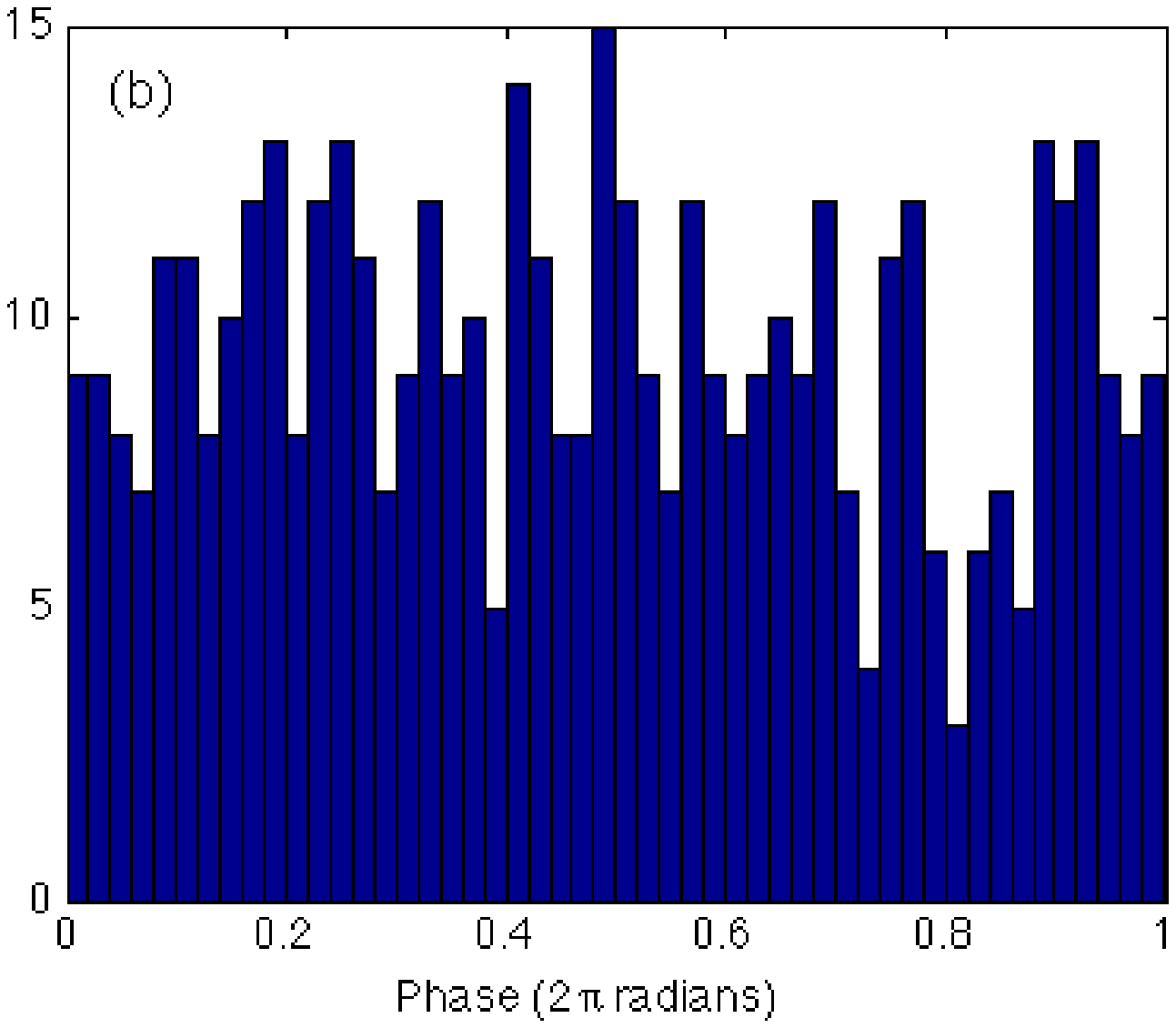}\\
 
    \includegraphics[width=0.23\textwidth,keepaspectratio=true]{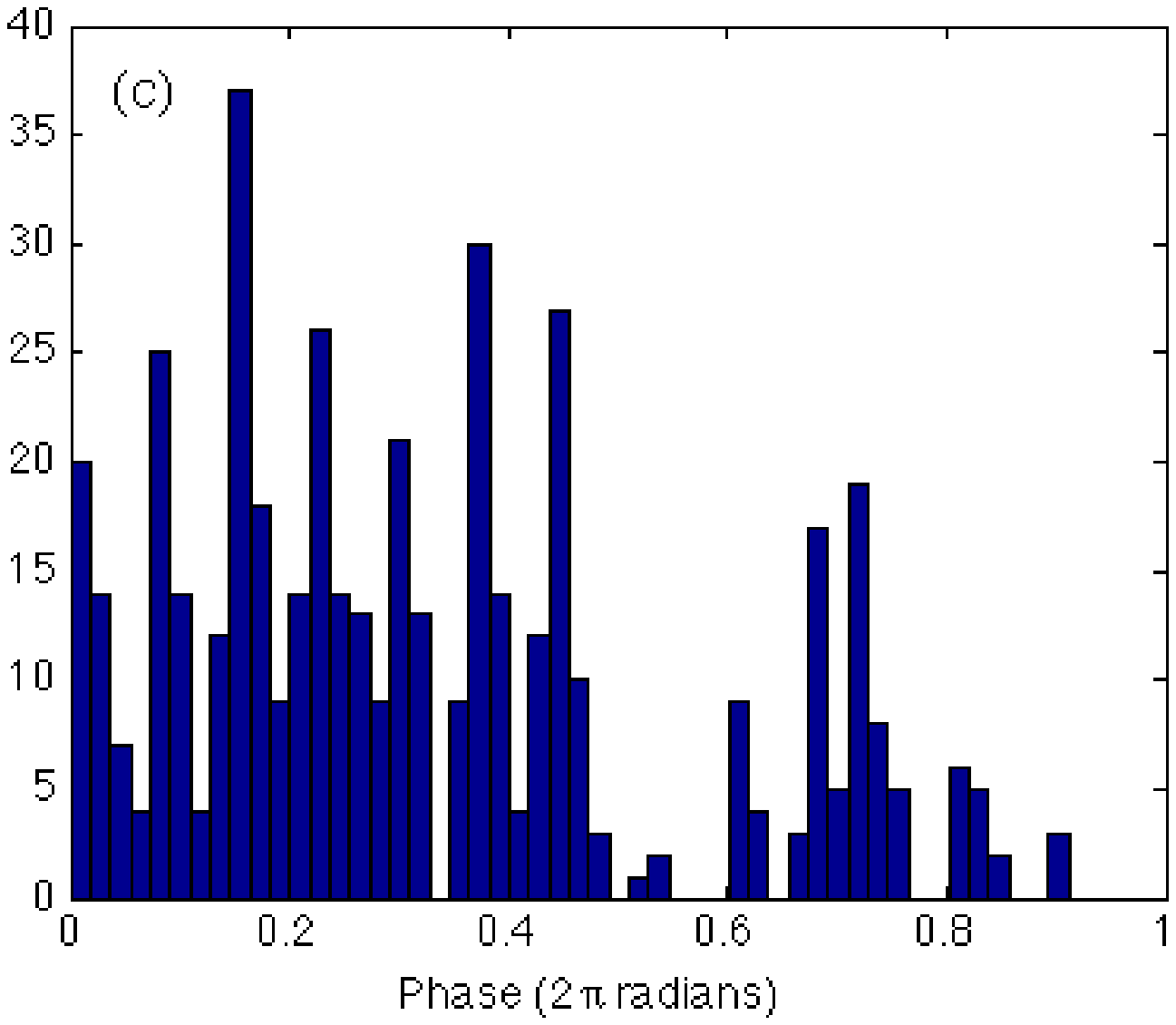}&

    \includegraphics[width=0.23\textwidth,keepaspectratio=true]{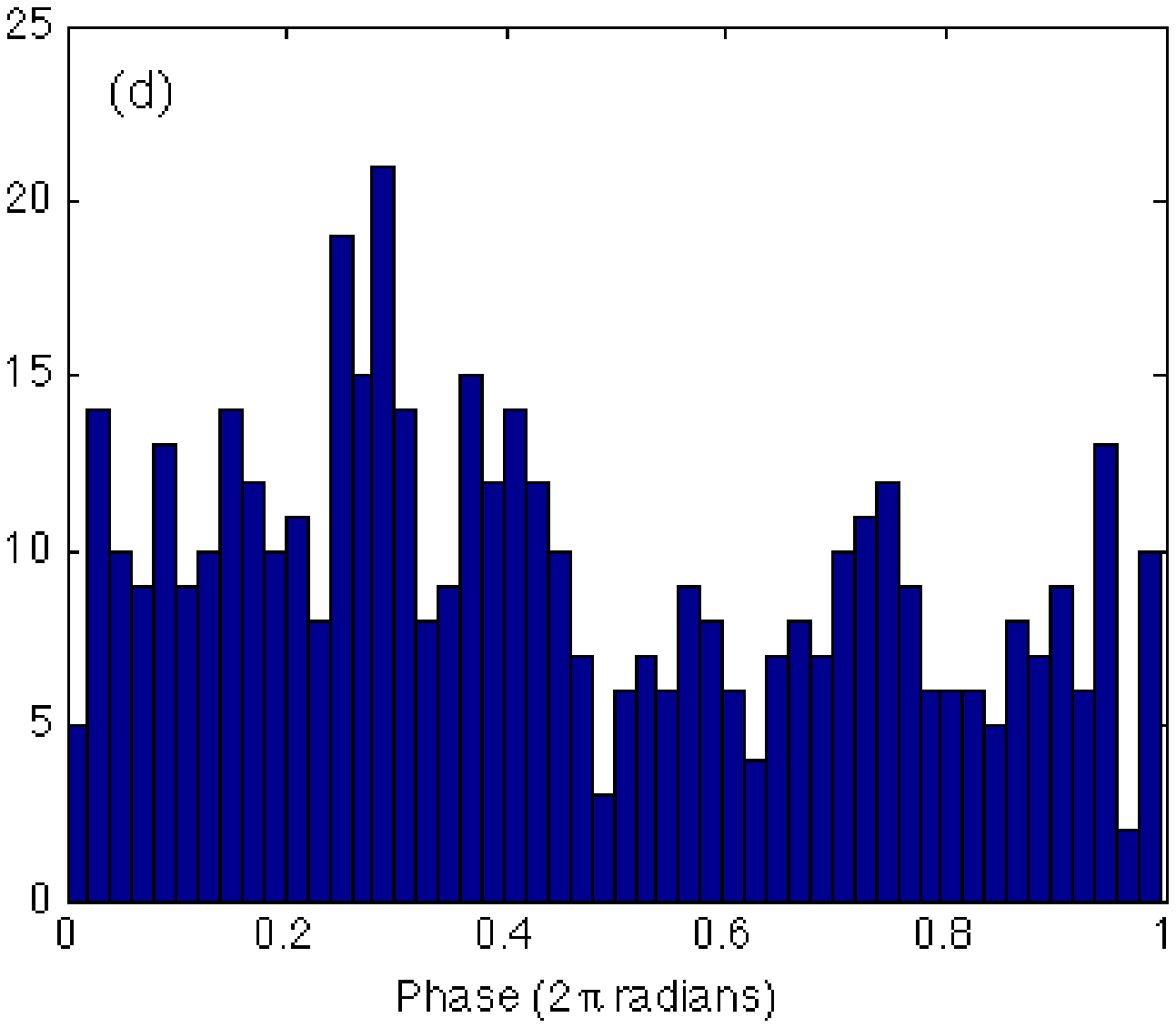}\\

\end{tabular}
\end{figure}

Using the frequencies from Table \ref{189631finalfreq} it was found that not all of them were sampled well across the whole phase of variation. In particular the smallest frequency
$f_3$ can be seen to be a little sparsely sampled (see Figure \ref{189631_phasesampling}). The paucity of data through phases of about $0.6$ and $0.9$ make confident 
model-fitting for mode identification more challenging.

\begin{figure}
  \centering
 \renewcommand{\tabcolsep}{1pt} 
  \caption[Line profile variation over pulsational phases for HD 189631]{Line profile variation phased to each determined frequency: (a) $f_1$, (b) $f_2$, (c) $f_3$, (d) $f_4$. The mean line profile has been subtracted in each case.}
\label{189631phase_colourplot}
  \begin{tabular}{cc}
     \includegraphics[width=0.23\textwidth,keepaspectratio=true] {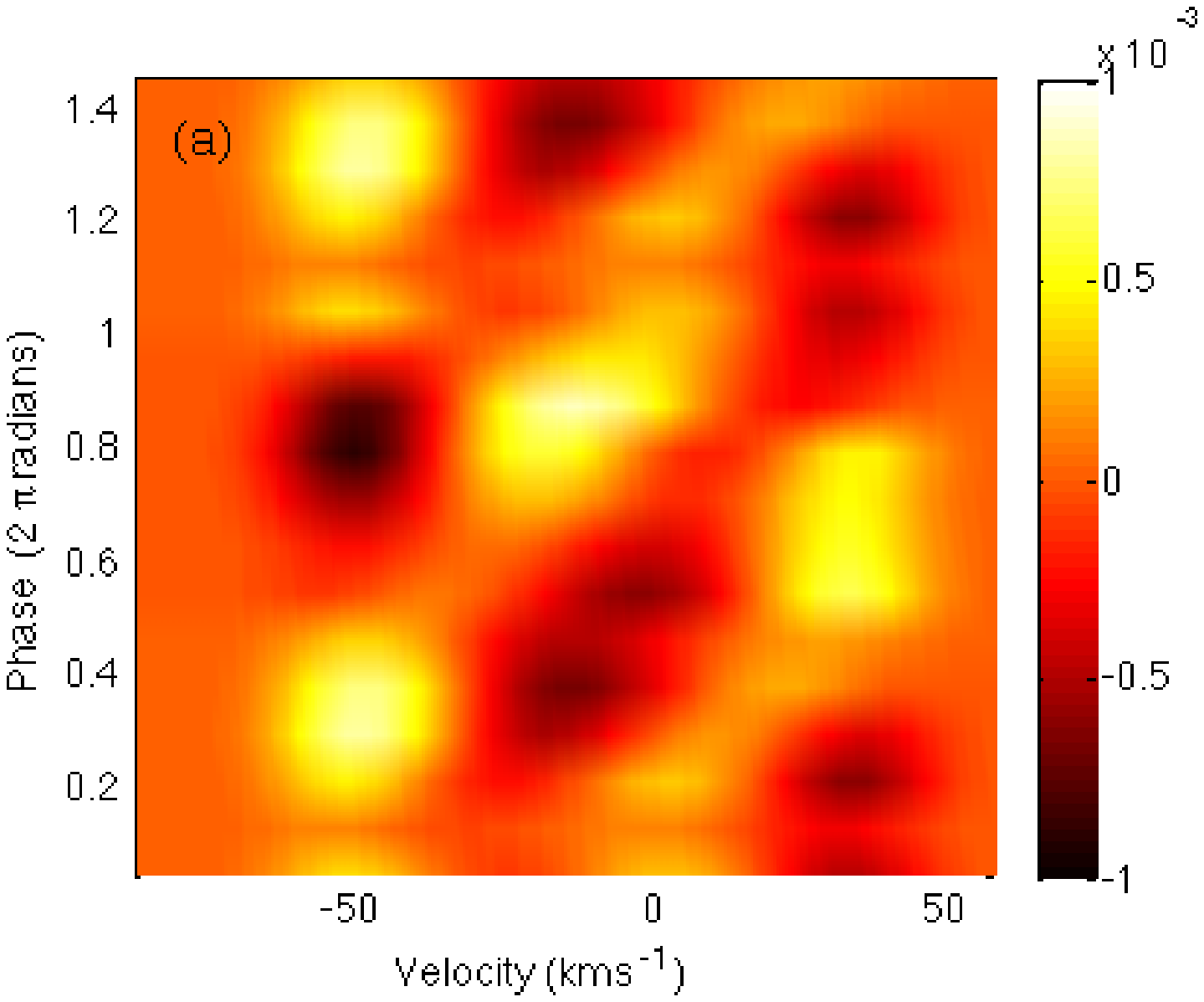}&
    \includegraphics[width=0.23\textwidth,keepaspectratio=true] {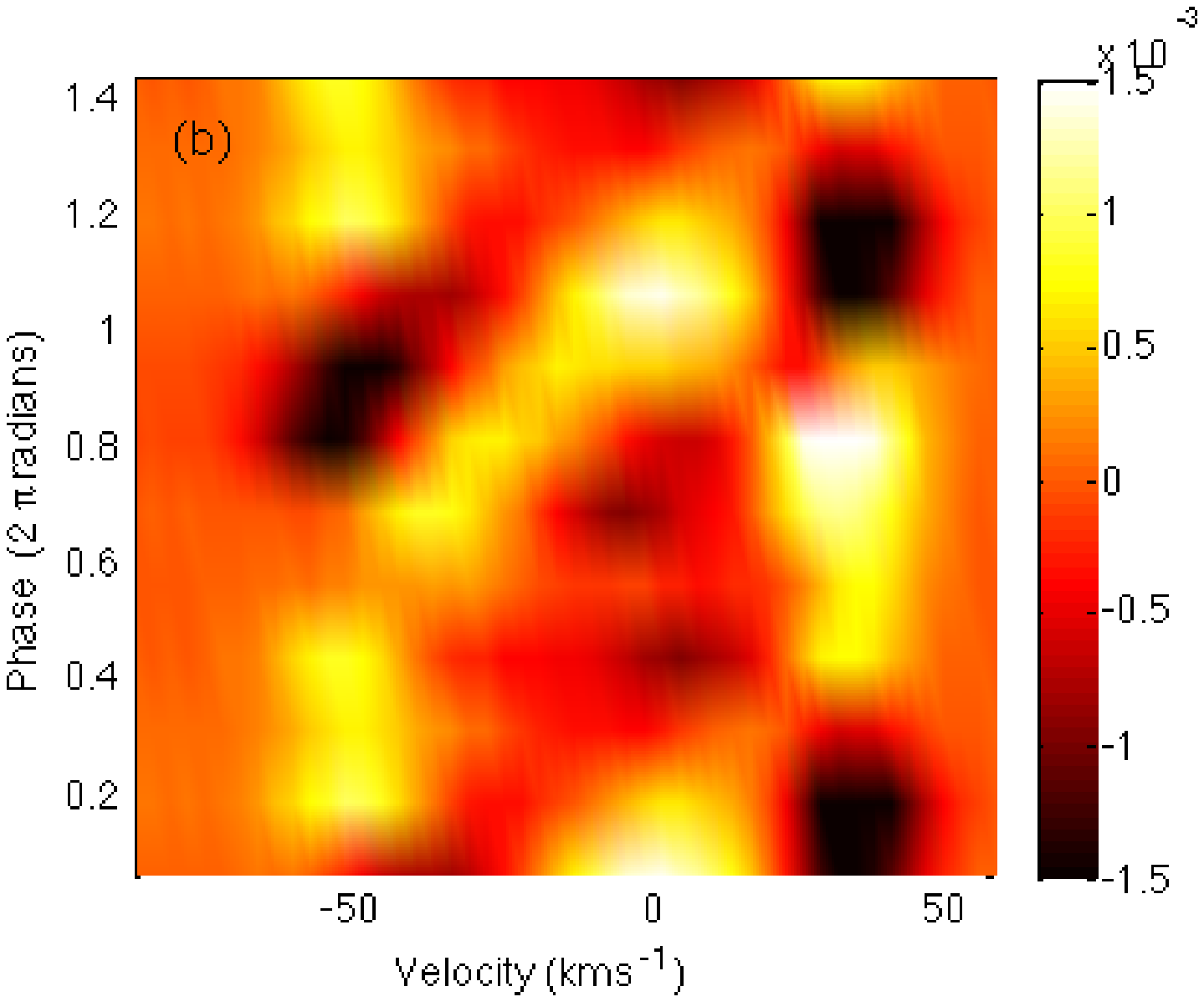}\\
   \includegraphics[width=0.23\textwidth,keepaspectratio=true] {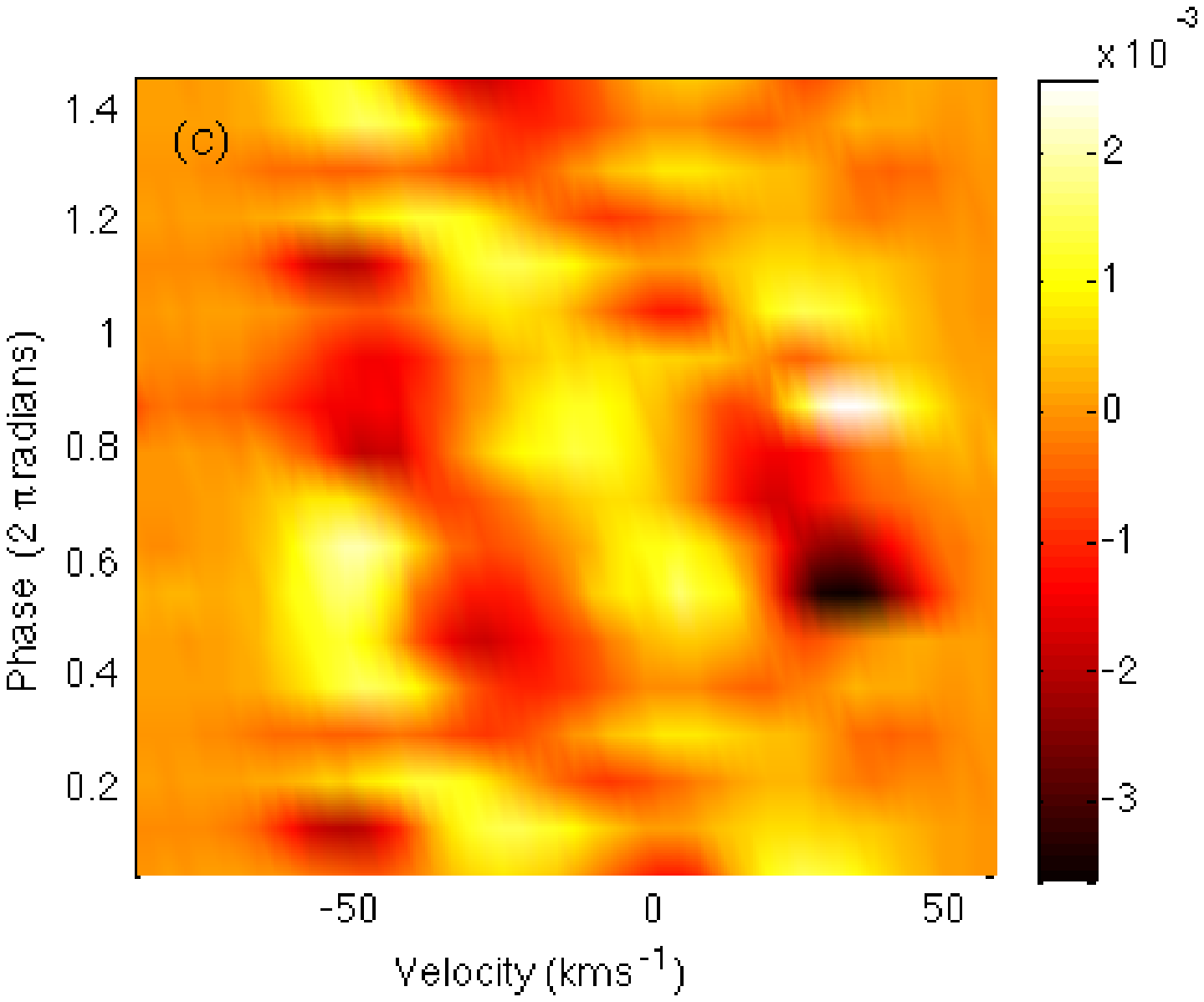}&
    \includegraphics[width=0.23\textwidth,keepaspectratio=true] {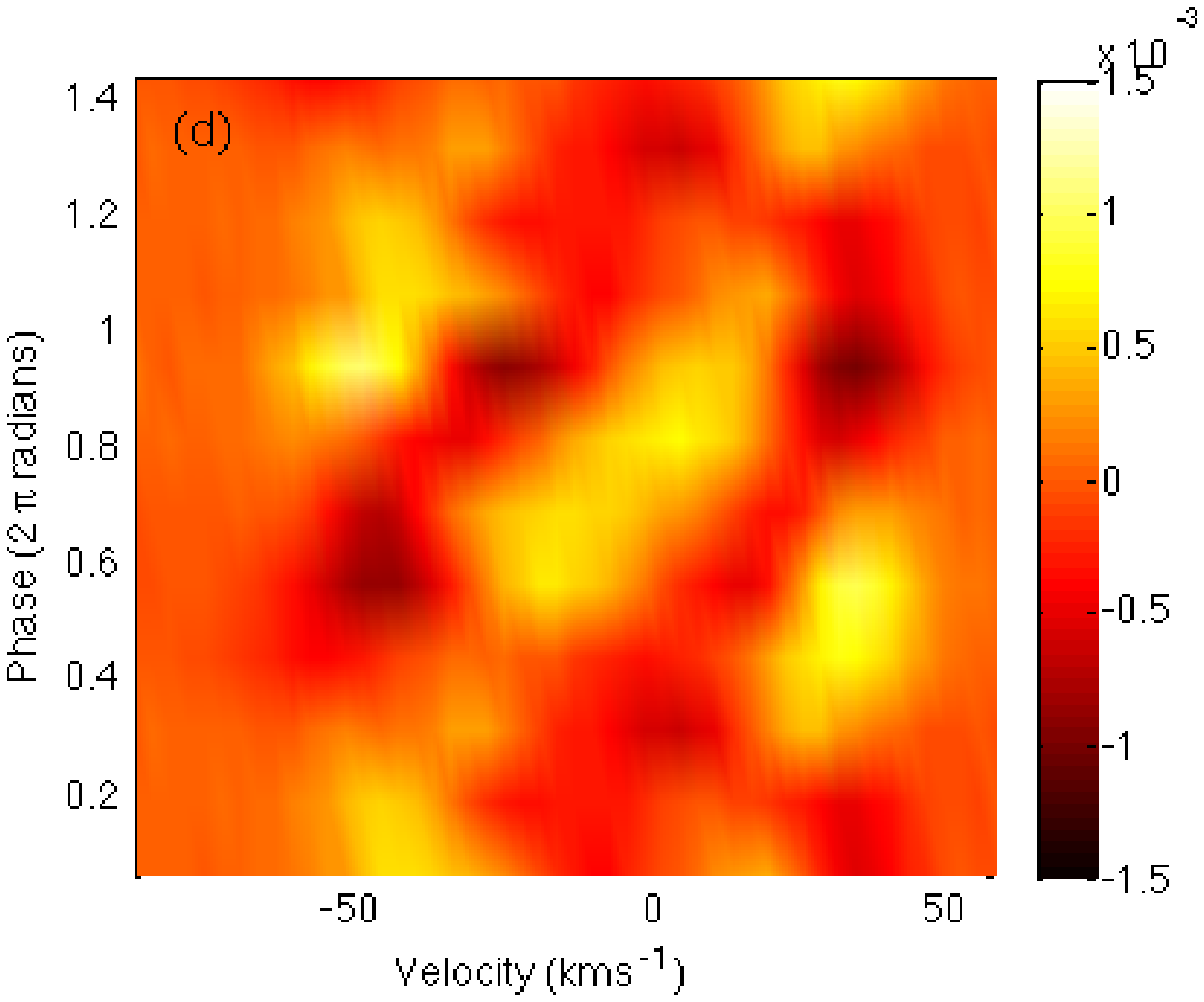}\\
\end{tabular}
\end{figure}

The variation of the line profiles over the phase of pulsation is shown in Figure \ref{189631phase_colourplot}. $f_2$ shows strong features in the wings that appear to remain stationary over 
some of the pulsational period. Also notable is the slight offset in $f_3$ (Figure \ref{189631phase_colourplot}c) which might be accounted for by poor sampling over the phase, skewing the mean profile away from a mean velocity of zero. The retrograde motion of $f_3$ across the line profile is clearly seen when compared to the other frequencies.

Aliases of these frequencies were also examined; $1$, $2$, and $3$ days $\pm f$, but these were much less strongly detected than those frequencies already identified.

\begin{table}
\caption[Finalised frequencies for HD 189631]{The finalised list of frequencies determined for HD 189631 from pixel-by-pixel analysis and from moments.}
\label{189631finalfreq}
\begin{minipage}{0.45\textwidth}
\centering
 \begin{tabular}{l|cc} 
  \hline \hline
  	&\multicolumn{2}{c}{Frequency}\\
	&d$^{-1}$	&	$\sigma(f)$\\
\hline
$f_1$	&	1.6774	&	0.0002	\\
$f_2$	&	1.4174	&	0.0002	\\
$f_3$	&	0.0714	&	0.0002	\\
$f_4$	&	1.8228	&	0.0002	\\
 \hline \hline
\end{tabular}
\end {minipage}
\end{table}

\section{Mode identification}
The first frequency ($f_1$ = $1.6774$ d$^{-1}$) was fit with models generated in \textsc{famias} using the parameters outlined in Table \ref{FAMIASparams189631}.
Care was taken to limit the lower bound of the inclination to always be above the inclination at which the equatorial velocity exceeded its critical upper limit.
For HD~189631 with an assumed mass of $1.5$ M$_{\odot}$ and radius of $1.6$ M$_{\odot}$ typical of $\gamma$~Doradus stars, \citep{1999PASP..111..840K}~
and a $v$sin$i$ of $43.6$~kms$^{-1}$, a critical inclination of~$6^{\circ}$~is found, above 
which equatorial rotation approaches a breakup scenario. The lower limit for the inclination therefore was set at $8^{\circ}$. \textsc{famias}~was permitted to examine combinations of parameters which would generate non-physical  scenarios, but these did not produce good fits

\begin{table}

\caption[FAMIAS parameters for HD 189631]{The parameters used for models generated for mode identification in HD 189631.}
\label{FAMIASparams189631}
\begin{minipage}{0.45\textwidth}
 \centering
 \begin{tabular}{l|rl}  
 \hline \hline
 
$R$&		1.5 --- 1.8 &R$_{\odot}$	\\
$M$&		1.4 --- 1.6 &M$_{\odot}$	\\
$T$&7500&K\\
log$g$	&4.2&\\
$[Fe/H]$&0.08&\\
Equivalent width&1.4 --- 1.9&kms$^{-1}$	\\
$v$sin$i$&40 --- 50&kms$^{-1}$\\
inclination& 8 --- 90&degrees\\
Intrinsic width&7 --- 10&kms$^{-1}$\\
Zero-point shift&-12 --- -8&kms$^{-1}$\\
$l$&0 --- 4&\\
$m$&-4 --- +4&\\
Amplitude&0.1 --- 5&kms$^{-1}$\\
Phase& 0 --- 1&$2\pi~$radians\\

\hline \hline
\end{tabular}\par
   \vspace{-0.75\skip\footins}
   \renewcommand{\footnoterule}{}
\end {minipage}
\end{table}

\begin{figure} 
 \includegraphics[width=0.45\textwidth,keepaspectratio=true]{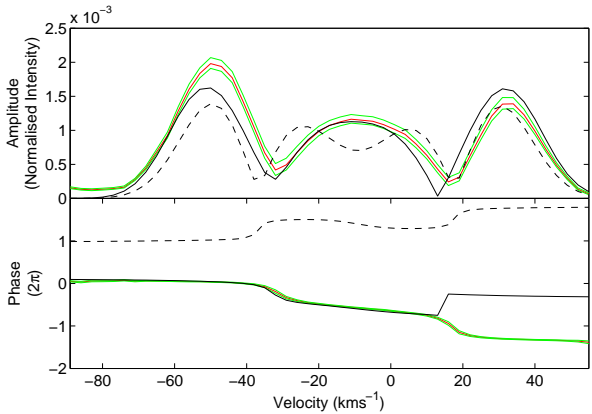}

 \caption[Mode identifications for $f_1$ of HD 189631]{Modes fitted to $f_1$ of HD 189631. Here the observed amplitude and phase across the line are given by the blue line, with error bounds in green. The best fit of the ($l$,$m$) = ($1,+1$) mode is given as the solid line and the best fit of the ($l$,$m$) = ($4, 0$) mode is given as the dashed line. Note that a difference of an integer in phase means that the pulsation has the same phase.}
 \label{189631_F1_fits}
\end{figure}

The minima of reduced
$\chi^2$ values for the models generated showed decisively that $f_1$ ($1.6774$~d$^{-1}$) was best fitted by an ($l,m$) = ($1, +1$) mode. This mode was fitted with a reduced $\chi^2$ of $11$, the next best fit (a $4, 0$ mode) had a reduced $\chi^2$ of $33$. This is reasonable considering the ratio of the rotation frequency to intrinsic pulsation frequency is $0.64$ ($\nu = 1.28$ in \citet{2003MNRAS.343..125T}). The fit of the phase is clearly much 
better for the best fit of the ($l$,$m$) = ($1,+1$) mode (Figure \ref{189631_F1_fits}). This identification 
agrees with that of \citet{2011MNRAS.415.2977M} for the dominant frequency. This best fit mode was achieved with an inclination of $56.0^{\circ}$.

\begin{figure}
 \centering
 \includegraphics[width=0.45\textwidth,keepaspectratio=true]{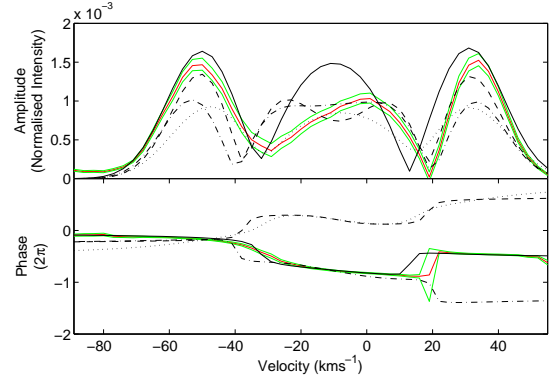}

 \caption[Mode identifications for $f_2$ of HD 189631]{Mode identifications for $f_2$ of HD 189631. Here the observed amplitude and phase across the line are given by the blue line, with error bounds in green. The ($l$,$m$) = ($1,+1$) mode is shown with a solid line, the ($l$,$m$) = ($4, 0$) mode is shown with a dashed line the ($l$,$m$) = ($2, -2$) mode is shown with the dash-dotted line, and the ($l$,$m$) = ($3, -2$) mode is shown with a dotted line.}
 \label{189631_f2_fits}
\end{figure}

The mode identification of $f_2$ = $1.4174$ d$^{-1}$ was similarly undertaken. It was best fit with an ($l,m$) = ($1, +1$) mode ($\chi^2 = 13$) but with an inclination of $75^{\circ}$. 
Other close fits were an ($l,m$) = ($2, -2$) (with an inclination of $13^{\circ}$ and a $\chi^2 = 24$) or an ($l,m$) = ($4, 0$) mode (with an inclination of $90^{\circ}$ and $\chi^2 = 25$). This difference in the 
inclinations of the best models is particularly significant when one considers that all modes appear in a star with only one rotational inclination angle (see Section \ref{189631combined}). The minima for the inclination are broad.
 The mode identified for this frequency by \citet{2011MNRAS.415.2977M} was a ($l,m$) = ($3, -2$) which we find to have a 
reduced $\chi^2$ value of 40, much greater than the best fit ($l,m$) = ($1, +1$) mode obtained here. Figure \ref{189631_f2_fits} shows the three best fits and that of \citet{2011MNRAS.415.2977M} for this frequency. The discontinuity in phase at around $20$ kms$^{-1}$~has 
large associated uncertainties and the phase is generally poorly fit by these models. The asymmetry of the central bump is also poorly modelled with \textsc{famias}.

\begin{figure}
 \includegraphics[width=0.45\textwidth,keepaspectratio=true]{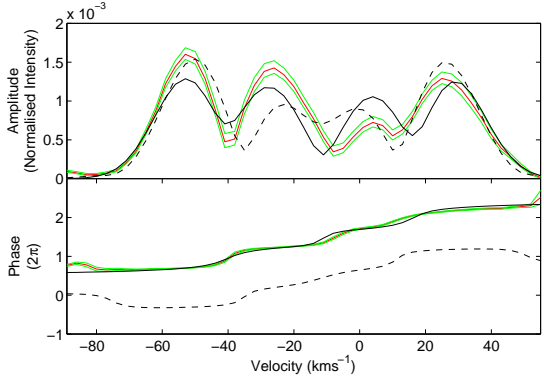}

 \caption[Mode identifications for $f_3$ of HD 189631]{Best fitted modes to $f_3$ of HD 189631. Here the observed amplitude and phase across the line are given by the blue line, with error bounds in green as before. The ($l$,$m$) = ($2, -2$) mode is shown with a solid line, the ($l$,$m$) = ($4, -2$) mode is shown with a dashed line.}
 \label{189631_f3_fits}
\end{figure}

The mode identification of $f_3$ = $0.0714$ d$^{-1}$ found two retrograde ($m<0$)~
  modes showing merit. It was best fit with an ($l,m$) = ($2, -2$) mode ($\chi^2 = 11$) with the other option being ($l,m$) = ($4, -2$) ($\chi^2 = 13$) mode. The difference in the amplitude 
profile from these two modes is small and, with the asymmetry in the central bump in the observed profile, they are similarly well fit by 
either a mode with a single central bump or a mode with two low-amplitude central bumps. The difference in the phase of pulsation across 
the line (Figure \ref{189631_f3_fits}) favours the ($l$,$m$) = ($2, -2$) mode. \citet{2011MNRAS.415.2977M} too found 
a ($l,m$) = ($2, -2$) mode for this frequency.

\begin{figure}
 \includegraphics[width=0.45\textwidth,keepaspectratio=true]{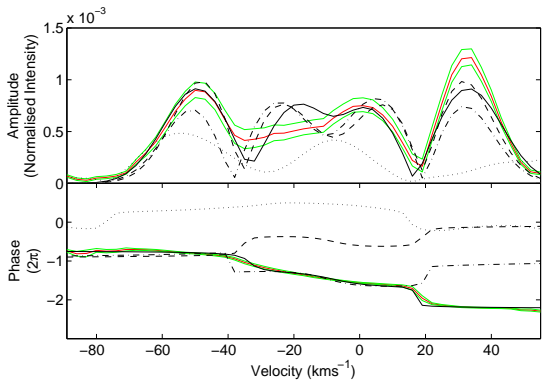}

 \caption[Mode identifications for $f_4$ of HD 189631] {Best fitted modes for $f_4$ in HD 189631. Here, as before, the observed amplitude and phase across the line are given by the blue line, with error bounds in green. The best fit of an ($l,m$) = ($1, +1$) mode is shown with a solid black line, the best fit of an ($l,m$) = ($2, +2$) is shown with a dash-dotted line, the best fit of an ($l,m$) = ($4, 0$) mode is shown with a dashed line, and the best fit of an ($l,m$) = ($4, +1$) mode is shown with a dotted line.}
 \label{189631_f4_fits}
\end{figure}

The mode identification for $f_4$ is much more uncertain than the prior modes. The amplitude is small, as it is found in the residuals from three stronger periodic variations.
\citet{2011MNRAS.415.2977M} found a ($l,m$) = ($4,+1$) mode but was uncertain and suggested a ($l,m$) = ($2,-2$) mode as an 
alternative. These prior results and recommendations are not in good agreement with those obtained here. There are good fits 
of the ($l,m$) = ($1,+1$) mode ($\chi^2 = 4$), followed by an ($l,m$) = ($2,+2$) ($\chi^2 = 11$) (Figure \ref{189631_f4_fits}), and more distantly followed by those of \citet{2011MNRAS.415.2977M}. In observing the phase of these 
pulsational models we see a poor fit to $f_4$ of the ($l,m$) = ($4,+1$) mode ($\chi^2 = 13$) as recommended by \citet{2011MNRAS.415.2977M}. 

\section{Combined mode identification} \label{189631combined}

Mode identification was undertaken with each frequency simultaneously, forcing the models to use one value of each stellar 
parameter for all modes being fitted. This assumes the pulsation axis for each mode is aligned with the rotation axis. Modes that 
are incompatible with each other generate a poor reduced $\chi^2$ when fitted together, such as 
those with very different optimal inclinations. It may be possible that pulsations are misaligned with each other or with the rotational 
axis but this has not been considered here.

All four frequencies were fit simultaneously with models of varying $l$ and $m$ values. As expected the
modes found from individual fitting were generally found again in the multimodal fit. The best fitting 
modes are shown in Table \ref{189631multimodal}.

\begin{table}

\caption{Reduced~$\chi^2$~values for multiple simultaneous mode identifications of HD 189631. Only the $l$, $m$ combinations with the lowest~$\chi^2$~values are shown.}
\label{189631multimodal}
\begin{minipage}{0.47\textwidth}
 \centering
 \begin{tabular}{c|rl|rl|rl|rl}  
  \hline \hline
		  &\multicolumn{2}{c}{$f_1$}	&\multicolumn{2}{c}{$f_2$}	&\multicolumn{2}{c}{$f_3$}	&\multicolumn{2}{c}{$f_4$}\\
Reduced~$\chi^2$ &$l,$	&$m$			&$l,$	&$m$			&$l,$	&$m$			&$l,$	&$m$		\\
 \hline
13	&	1,	&	1  	&	1,	&	1  	&	2,	&	-2  	&	1,	&	1  	\\
18	&	1,	&	1  	&	2,	&	-2  	&	2,	&	-2  	&	1,	&	1  	\\
20	&	1,	&	1  	&	1,	&	1  	&	2,	&	-2  	&	0,	&	0  	\\
20	&	1,	&	1  	&	1,	&	1  	&	2,	&	-2  	&	1,	&	0  	\\
20	&	1,	&	1  	&	1,	&	1  	&	2,	&	-2  	&	3,	&	0  	\\
20	&	1,	&	1  	&	1,	&	1  	&	2,	&	-2  	&	4,	&	1  	\\
20	&	1,	&	1  	&	3,	&	-2  	&	2,	&	-2  	&	1,	&	1  	\\
 \hline \hline
\end{tabular}
\end {minipage}
\end{table}

\section{Discussion}

The best fit for each frequency was that of $f_1$ being an ($l$,$m$) = ($1, +1$) mode, $f_2$ an ($l$,$m$) = ($1, +1$) 
mode, $f_3$ an ($l, m$) = ($2, -2$) mode and $f_4$ an ($l$,$m$) = ($1, +1$) mode. 

Modes identified from simultaneous fits of all frequencies are in agreement with those obtained by fitting modes to frequencies individually, but are conspicuously not in complete agreement with those of \citet{2011MNRAS.415.2977M}. The modes identified here are however in excellent agreement with those of \citet{Tkachenko2013}, who used a modified least squares deconvolution method to form the line profiles where a more simple cross-correlation was applied here. That the same modes and frequencies were identified using different methods and data sets further strengthens the confidence one may place in these results and those of \citet{Tkachenko2013}. 
 
 An aspect to note is the difference in the $\chi^2$ values obtained here 
 versus \citet{2011MNRAS.415.2977M}. \citet{2011MNRAS.415.2977M} fit not just the phase and standard deviation profile but the line profile too.
The much larger scale in the line profile means that any deviation from the fitted model to the observed line profile dominates the $\chi^2$ obtained at the expense of the fits of the deviation and phase profiles.
  
\begin{figure}
  \begin{minipage}{0.47\textwidth}

 \includegraphics[width=\textwidth,keepaspectratio=true]{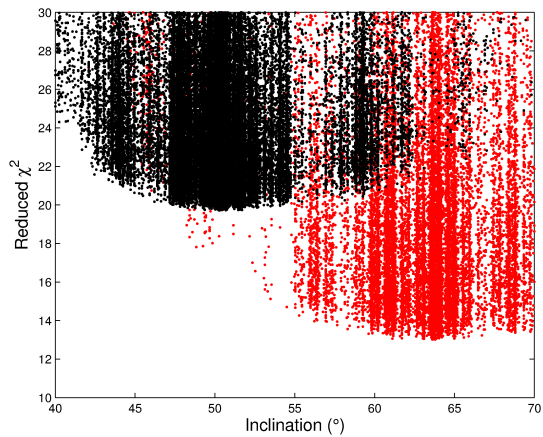}

 \caption{Reduced $\chi^2$ values for different inclinations used in the model generation for the fitting to $f_2$ of an ($l$,$m$) = ($1, +1$) mode in red and ($l$,$m$) = ($3, -2$) mode in black (as found by \citet{2011MNRAS.415.2977M}). These fits were obtained from the simultaneous multi-modal fitting of all four frequencies.}
 \label{189631inclination}
 \end{minipage}
\end{figure}

A possible explanation of this discrepancy in these mode identifications is that more data is available for this study. Improvements were acheived in the frequencies determined in this study, and similarly in the density of data over the narrow transitions seen in the phase for each frequency allowing closer modelling. Alternatively perhaps the presence of multiple local $\chi^2$ minima 
 in the parameter-space searched led to to the different modes identified. This is problematic where parameters with the possibility of multiple minima, like the azimuthal order $m$ are concerned. The broadness of the minima in Figure \ref{189631inclination} gives some idea of the low level of sensitivity to the 
inclination parameter $i$. The inclination $i$, allows for viewing of different sums over the pulsational vector field in the line-of-sight. In the ($l$,$m$) = ($3, -2$) case it determines 
how the equatorial node is presented. What should be noted here is that for low ratios of vertical to horizontal displacement as found in g-mode pulsations, \textsc{famias} does not treat $i$ well. A 
more robust determination of $i$ could be used.

A noteworthy result of this mode identification is the detection of  ($l$,$m$) = ($1, +1$) modes for three of the frequencies. There may well be some selection effect at 
work either observationally  due to the strong driving \citep{2011MNRAS.417..591B} and resultant large amplitudes \citep{2012MNRAS.427.2512B} for this modal geometry. 

HD~189631 is typical of $\gamma$~Doradus stars except for the detection of the retrograde pulsation $f_3$. There have been very few of these identified spectroscopically to date. It may simply be that the similarity between the stellar rotational frequency and the pulsational frequency makes them more challenging to detect as the scenario approaches that of a standing wave as we observe it.

\section{Acknowledgements}
This work was supported by the Marsden Fund.

The authors acknowledge the assistance of staff at
Mt John University Observatory, a research station of the University of
Canterbury.

This research has made use
of the \textsc{SIMBAD} astronomical database operated at the CDS in
Strasbourg, France.

Mode identification results obtained with the software package \textsc{FAMIAS} developed
in the framework of the FP6 European Coordination action HELAS
(http://www.helas-eu.org/).

\bibliography{references}{}
\bibliographystyle{apj}

\end{document}